\documentclass[reprint,longbibliography,aip]{revtex4-2}

\usepackage{graphicx} 
\usepackage{amsmath}
\usepackage{ amssymb }

\usepackage{color}

\usepackage{float} 
\usepackage{wrapfig} 

\usepackage[makeroom]{cancel} 
\usepackage{comment}

\usepackage{lipsum} 

\usepackage{outlines}
\usepackage{amsmath}

\DeclareMathOperator\arccosh{arccosh}

\graphicspath{{./figs/}} 

\newcommand{\beq}{\begin{equation}}
\newcommand{\eeq}{\end{equation}}

\newcommand{\bvec}{\begin{pmatrix}}
\newcommand{\evec}{\end{pmatrix}}
\newcommand{\lp}{\left(}
\newcommand{\rp}{\right)}
\newcommand{\pa}[2]{\frac{\partial #1}{\partial #2}}

\newcommand{\llangle}{\left \langle}
\newcommand{\rrangle}{\right \rangle}

\newcommand{\ve}[1]{\mathbf{#1}}

\newcommand{\Gammav}{\ve{\Gamma}}
\newcommand{\Dv}{\ve{D}}

\AtBeginDocument{%
\newwrite\bibnotes
\def\bibnotesext{Notes.bib}
\immediate\openout\bibnotes=\jobname\bibnotesext
\immediate\write\bibnotes{@CONTROL{REVTEX41Control}}
\immediate\write\bibnotes{@CONTROL{%
		apsrev42Control,author="08",editor="1",pages="1",title="0",year="1"}}
\if@filesw
\immediate\write\@auxout{\string\citation{apsrev42Control}}%
\fi
}%

\begin{document}



\title{Electron Tail Suppression and Effective Collisionality due to Synchrotron Emission and Absorption in Mildly Relativistic Plasmas}

\author{Ian E. Ochs}
\email{iochs@princeton.edu}
\affiliation{Department of Astrophysical Sciences, Princeton University, Princeton, New Jersey 08540, USA}

\author{Mikhail E. Mlodik}
\affiliation{Department of Astrophysical Sciences, Princeton University, Princeton, New Jersey 08540, USA}

\author{Nathaniel J. Fisch}
\affiliation{Department of Astrophysical Sciences, Princeton University, Princeton, New Jersey 08540, USA}

\date{\today}

\begin{abstract}
	Synchrotron radiation losses are a significant cause of concern for high-temperature aneutronic fusion reactions such as proton-Boron 11.
	The fact that radiation losses occur primarily in the high-energy tail, where the radiation itself has a substantial impact on the electron distribution, necessitates a self-consistent approach to modeling the diffusion and drag induced by synchrotron absorption and emission.
	Furthermore, an accurate model must account for the fact that the radiation emission spectrum is momentum-dependent, and the plasma opacity is frequency-dependent.
	Here, we present a simple Fokker-Planck operator, built on a newly-solved-for blackbody synchrotron diffusion operator, which captures all relevant features of the synchrotron radiation.
	Focusing on magnetic mirror fusion plasmas, we show that significant suppression of the electron distribution occurs for relativistic values of the perpendicular electron momentum, which therefore emit much less radiation than predicted under the assumption of a Maxwell-J\"uttner distribution.
\end{abstract}

\maketitle


\section{Introduction}

In the 1950s, electron cyclotron radiation was a cause of great concern to the fusion energy community, because an electron in a magnetic field radiates its energy away on a timescale $\tau_R \sim (\text{3.0 s})/B_T^2$, where $B_T$ is the magnetic field strength in Tesla.
Extrapolating to a modern 10 T reactor, one finds a confinement time of 1/30s, much too short to meet the Lawson criterion.
However, seminal work by Trubnikov and Kudryatsev in the USSR\cite{Trubnikov1958PlasmaRadiation} and by Drummond and Rosenbluth in the USA\cite{Drummond1963CyclotronRadiation} showed that most of the radiation is immediately reabsorbed by the plasma.
Thus, study of electron cyclotron radiation became mainly useful as a plasma diagnostic, with a focus on the emission, absorption, and propagation properties of the radiation given fixed plasma conditions;\cite{Costley1974ElectronCyclotron,Bornatici1983ECR,Costley200950Years}
an approach shared even by studies concerned with the radiative losses.\cite{Albajar2007ElectronCyclotron,Kukushkin2008ECR} 
Study of the effects of the waves themselves on the kinetic electron distribution became focused primarily on the narrow bands of electron cyclotron waves deliberately launched by wave antennae for heating and current drive.\cite{fisch1980creating,Prater2004HeatingCurrent}
Meanwhile, likely because of its lack of importance to DT fusion, there have been few studies\cite{Bornatici1994RelativisticFokkerPlanck} of the effect of a thermal electron cyclotron radiation on the electron distribution itself.
Where such effects are analyzed, focus is generally reserved for the fastest electrons, such as runaways, for which the plasma has almost exclusively been treated as optically thin, considering emission of radiation (radiative drag) but not absorption (radiative diffusion).\cite{Harvey1992CQL3DFokkerplanck,Boozer2015TheoryRunaway,Carbajal2017SynchrotronEmission,Breizman2019PhysicsRunaway}

Deuterium-tritium is not the only possible fusion reaction.
It is, however, the easiest, occurring with a large cross section at relatively low ion energies.
Nevertheless, the drawbacks involved with tritium handling and fast neutron production have led to a resurgence of interest in aneutronic fusion schemes.
The p-B11 reaction, in particular, is the focus of substantial study,\cite{Volosov2006ACT,Volosov2011Problems,Hay2015Ignition,Putvinski2019,Ochs2022ImprovingFeasibility,kolmes2022waveSupported,Ochs2024LoweringReactor,Magee2019DirectObservation,Eliezer2020NovelFusion,Ruhl2022LaserPB11,Istokskaia2023MultiMeVAlpha,Magee2023FirstMeasurements}, partly due to the fact that its cross section \cite{Sikora2016CrossSection} was recently found to be larger than previously thought. \cite{Rider1995,Nevins2000CrossSection}

Because p-B11 fusion takes place at much higher temperatures---typically 300 keV for ions, and 160 keV for electrons\cite{Putvinski2019,Ochs2022ImprovingFeasibility,kolmes2022waveSupported,Ochs2024LoweringReactor}---the electrons are much more relativistic than in a DT plasma.
As a result, the electron cyclotron radiation occurs at higher harmonics, which are less efficiently reabsorbed by the plasma.
Indeed, the qualitative differences in the spectrum often lead it to be referred to by a different name: synchrotron radiation.

Recently, it was shown that because (i) higher energy (more relativistic) electrons radiate at higher harmonics, and (ii) the radiation emitted at higher harmonics escapes the plasma more easily, the total energy radiated away by an aneutronic fusion plasma is heavily dependent on the tail electrons.\cite{Mlodik2023SensitivitySynchrotron}
Indeed, if the tail of the distribution is cut off, the radiation losses can be reduced by an order of magnitude.
Lesser, but still pronounced, tail sensitivity occurs for relativistic bremsstrahlung radiation.\cite{Munirov2023SuppressionBremsstrahlung}
Thus, accurate modeling of the tail distribution becomes extremely important for calculations of radiative loss.

At the same time as the spectrum changes, the synchrotron radiation also becomes relatively stronger, as the Coulomb collisions grow weaker.
The relative strength of synchrotron to Coulomb collisions is roughly determined by the parameter\cite{Bornatici1994RelativisticFokkerPlanck} $\epsilon_S \gtrsim 1$, where:
\begin{align}
	\epsilon_S \equiv \frac{2}{3 } \frac{\Omega_0^2}{\omega_{pe}^2} \frac{u^3}{\lambda_{ee} \gamma}, \label{eq:BornaticiParameter}
\end{align}
and where $u = p/mc$ is the normalized electron momentum, $\gamma = \sqrt{1+u^2}$ is the Lorentz factor, $\lambda_{ee}$ is the Coulomb logarithm, $\Omega_0 = e B / m_e c$ is the (non-relativistic) gyrofrequency, and $\omega_{pe} = 4\pi e^2 n_e/m_e$ is the electron plasma frequency.
For a 10T magnetic field and an electron density of $1.6 \times 10^{14}$ cm$^{-3}$, the second fraction is approximately 6, and so synchrotron effects become important around $u \sim 1$, i.e. as the plasma becomes mildly relativistic.
Thus, in the electron tails that largely determine the radiation losses, the electron distribution itself is likely to be largely shaped by the radiation.

To accurately model radiation losses, and indeed any kinetic processes in the electron tail, it is necessary to have an operator that accurately models the key effects of synchrotron emission and absorption, including the effects of momentum-dependent emission spectra and frequency-dependent plasma opacities.
The goal of the current paper is to provide just such an operator, with a minimum of computational complexity, and to demonstrate the key behaviors that result from the inclusion of synchrotron effects on aneutronic-fusion-relevant kinetic electron distributions.

To develop the requisite operator, our starting point is Bornatici's\cite{Bornatici1994RelativisticFokkerPlanck} diffusion operator for a general radiation spectrum.
The first step is to evaluate this diffusion operator for the specific case a blackbody spectrum of radiation.
Although Ref.~\onlinecite{Bornatici1994RelativisticFokkerPlanck} derived several important properties of diffusion operator for a blackbody spectrum---in particular showing that it respected the Einstein relation in thermal equilibrium---it stopped short of deriving the full operator. 
We complete this derivation in Sec.~\ref{sec:BlackbodyDiffusion}, showing that the blackbody operator consists of two parts: (i) a thermalizing component that enforces thermal equilibrium along the path of radiative drag, and (ii) a smaller pitch-angle scattering component, mostly relevant at marginal relativistic momentum.

The blackbody diffusion operator, of course, only applies to the case of an optically thick plasma.
However, in Sec.~\ref{sec:DiffusionMixedOpacity}, we show how it can be generalized to plasmas with frequency-dependent opacity, by scaling the blackbody diffusion operator by the fraction $\lambda(\ve{u})$ of radiation emitted by an electron with momentum $\ve{u}$ that escapes the plasma.
This operator obeys a sensibly generalized Einstein relation in the presence of radiative losses, thus respecting local energy conservation in phase space while reducing to correct optically-thick and optically-thin limits.
In Section~\ref{sec:OpacityFunctions}, we show how to calculate $\lambda(\ve{u})$, which in general requires also calculating an effective cutoff frequency $\omega^*$ above which the plasma becomes optically thin.
The two functions $\lambda$ and $\omega^*$ we term the ``opacity functions'' for the synchrotron radiation.
Section~\ref{sec:Summary} summarizes the model.

With the mixed-opacity operator thus derived, we turn to look at the resulting electron distributions.
In Sec.~\ref{sec:QuasiSteadyState}, we examine the quasi-steady states of the synchrotron operator, working alone and with fixed opacity functions. We show that the portion of the distribution with $u_\perp\gtrsim 2$ tends to be strongly suppressed for laboratory-scale plasmas.
In Sec.~\ref{sec:MirrorConfinement}, we examine how synchrotron radiation and collisions combine to shape the electron distribution and confinement time in a magnetic mirror confinement scheme.
We find that the significant suppression in perpendicular momentum tends to produce a more isotropic distribution than would be generally predicted for a mirror plasma in the absence of synchrotron radiation.
We also find that, while the synchrotron radiation can in general either increase or decrease the electron particle confinement time, for the most reactor-relevant scenarios, it tends to improve electron confinement substantially.

Finally, in Sec.~\ref{sec:Conclusion}, we discuss how to incorporate this work into fusion power balance calculations, and discuss the possible impacts of simplifying assumptions.
We also discuss how the operator created here could be relevant in ultrarelativistic astrophysical scenarios, given recent work on the importance of synchrotron drag in destabilizing a broad swathe of kinetic distributions.\cite{Bilbao2023RadiationReaction,Bilbao2024RingMomentum,Zhdankin2023SynchrotronFirehose}

\section{Synchrotron Diffusion Operator for Blackbody Spectrum} \label{sec:BlackbodyDiffusion}

In this section, we will derive the diffusion operator for electron-cyclotron and synchrotron radiation in the case of a blackbody spectrum of radiation.
Our starting point is
Bornatici's diffusion and drag operators\cite{Bornatici1994RelativisticFokkerPlanck} for a general spectrum of synchrotron radiation, assuming a tenuous\cite{Bornatici1983ECR,Mlodik2023SensitivitySynchrotron} plasma:
\begin{align}
	\pa{f}{t} &= \frac{1}{\sqrt{g}}\pa{}{\ve{u}} \cdot \left[ \sqrt{g} \left(- \ve{\Gamma} f + \ve{D} \cdot \pa{f}{\ve{u}} \right) \right] \label{eq:FundamentalDiffusion}\\
	\ve{\Gamma} &= -\frac{1}{mc^2} \sum_{i,s} 2\pi \int_0^\infty d\omega \int_{-1}^{1} d \cos \theta \eta_s^{(i)} \ve{d^{(s)}}\\
	\ve{D} &= \frac{1}{mc^2} \sum_{i,s} 2\pi \int_0^\infty d\omega \int_{-1}^{1} d \cos \theta \frac{(2 \pi)^3 I^{(i)}}{m \omega^2} \eta_s^{(i)} \ve{d^{(s)}} \ve{d^{(s)}}, \label{eq:DiffusionBornatici}
\end{align}
where
\begin{align}
	\ve{u} &\equiv (u_\perp,u_\parallel) \equiv \frac{1}{mc^2}(p_\perp,p_\parallel)\\
	\ve{d^{(s)}} &= \left(\frac{s \Omega_0}{\omega} \frac{1}{u_\perp},\cos{\theta}\right)\\
	\sqrt{g} &= 2 \pi u_\perp.
\end{align}
Here, $\Omega_0=e B/m_e c$ is the (nonrelativistic) cyclotron frequency, $\eta_s^{(i)}(\omega,\theta)$ is the $s$-harmonic emissivity, and $I^{(i)}(\omega,\theta)$ is the intensity of radiation.
It should be noted that $\eta_s^{(i)}$ contains a $\delta$ function: 
\begin{align}
	\eta_s^{(i)} \propto \delta\left(\gamma - u_\parallel \cos \theta - \frac{s \Omega_0}{\omega} \right), \label{eq:DeltaFunction}
\end{align}
where $\gamma^2 = 1+u^2$, which sets the emission angle for the radiation a the $s^\text{th}$ harmonic:
\begin{align}
	\cos \theta = \frac{1}{u_\parallel} \left(\gamma - \frac{s \Omega_0}{\omega}\right).
\end{align}

For a blackbody spectrum in the Rayleigh-Jeans limit:
\begin{align}
	\frac{(2 \pi)^3 I^{(i)}}{m \omega^2} = \frac{T_{bb}}{mc^2},
\end{align}
implying that:
\begin{align}
	\ve{D}_\text{bb} &= \frac{T_{bb}}{m^2c^4} \sum_{i,s} 2\pi \int_0^\infty d\omega \int_{-1}^{1} d \cos \theta \eta_s^{(i)} \ve{d^{(s)}} \ve{d^{(s)}}. \label{eq:DiffusionBornaticiBB}
\end{align}

Thus, we have:
\begin{align}
	\ve{\Gamma} &= -\frac{1}{mc^2}  \bvec \frac{U_1}{u_\perp}\\ \frac{\gamma U_0 - U_1}{U_\parallel} \evec \\
	\ve{D}_\text{bb} &= \frac{T_{bb}}{m^2c^4} \bvec \frac{U_2}{u_\perp^2} & \frac{\gamma U_1 - U_2}{u_\parallel u_\perp}\\ \frac{\gamma U_1 - U_2}{u_\parallel u_\perp} & \frac{\gamma^2 U_0 - 2\gamma U_1 + U_2}{u_\parallel^2}\evec 
\end{align}
where
\begin{align}
	U_m = \sum_{i,s} 2\pi \int_0^\infty d\omega \int_{-1}^{1} d \cos \theta \left(\frac{s \Omega_0}{\omega}\right)^m\eta_s^{(i)}.
\end{align}

Ref.~\onlinecite{Bornatici1994RelativisticFokkerPlanck}'s Eq.~(21) gives results for $U_0$ and $U_1$, citing Ref.~\onlinecite{Melrose1980PlasmaAstrophysics}:
\begin{align}
	U_0 &= m c^2 \nu_{R0} u_\perp^2; \quad U_1 = m c^2 \nu_{R0} u_\perp^2 \frac{\gamma_\perp^2}{\gamma},
\end{align}
where
\begin{align}
	\nu_{R0} &= \frac{2}{3} \frac{e^2 \Omega_0^2}{m c^3}; \quad \gamma_\perp^2 = 1 + u_\perp^2. \label{eq:nuR0}
\end{align}
This leads to Ref.~\onlinecite{Bornatici1994RelativisticFokkerPlanck}'s Eq.~(22):
\begin{align}
	\Gammav = -\nu_{R0} \frac{\gamma_\perp^2}{\gamma} \ve{g}; \quad \ve{g} \equiv \bvec u_\perp \\ \frac{u_\perp^2}{\gamma_\perp^2} u_\parallel \evec \label{eq:GammaBornatici}
\end{align}
It can be seen that this radiative drag occurs along $\hat{u}_\perp$ for nonrelativistic particles, and along $\ve{u}$ for highly relativistic particles.
It can also be seen that the synchrotron radiation is associated with an effective collision timescale:\cite{Bernstein1981RelativisticTheory,Bornatici1994RelativisticFokkerPlanck}
\begin{align}
	\nu_R &\equiv \nu_{R0} \frac{\gamma_\perp^2}{\gamma} .
\end{align}
However, Ref.~\onlinecite{Bornatici1994RelativisticFokkerPlanck} never calculated $U_2$, which requires a much less friendly integration.

Avoiding the complexity for now of actually calculating $U_2$, let us write:
\begin{align}
	U_2 = m c^2 \nu_{R0} u_\perp^2 \lp \frac{\gamma_\perp^2}{\gamma} \rp^2 \lp 1+\frac{u_\parallel^2}{u_\perp^2} \Delta \rp. \label{eq:U2FromDelta}
\end{align}
Note that, since $\Delta$ is arbitrary, this is completely general.
Then, after a bit of algebra, $\ve{D}$ can be written as:
\begin{align}
	\ve{D}_\text{bb} = \nu_{R0} \frac{T_{bb}}{m c^2} \lp \frac{\gamma_\perp^2}{\gamma}\rp^2 \frac{1}{u_\perp^2} \left(\ve{G} + \Delta \ve{H} \right), \label{eq:DiffusionMatrixDecomposition}
\end{align}
where
\begin{alignat}{2}
	\ve{G} &= \bvec u_\perp^2 & \frac{u_\perp^3 u_\parallel}{\gamma_\perp^2}\\ \frac{u_\perp^3 u_\parallel}{\gamma_\perp^2} & \frac{u_\perp^4 u_\parallel^2}{\gamma_\perp^4}\evec = \ve{g} \ve{g}\\
	\ve{H} &= \bvec u_\parallel^2 & -u_\perp u_\parallel\\ -u_\perp u_\parallel & u_\perp^2 \evec = \ve{h} \ve{h}, \quad &\ve{h} \equiv \bvec u_\parallel \\ -u_\perp \evec
\end{alignat}

Now, we will show that the roles of the $\ve{G}$ matrix is to enforce the Einstein diffusion-drag relation via 1D diffusion along the radiative drag path, and the role of the $\ve{H}$ matrix is to provide pitch-angle scattering (providing, of course, that $\Delta > 0$).
Thus, just from the structure of the diffusion matrix and the values of $U_0$ and $U_1$, we can show that the operator respects important conservation relations without even knowing the value of $U_2$ (or $\Delta$).

\subsection{Pitch Angle Scattering from $\ve{H}$}

A diffusion matrix $\ve{D}$ represents pitch angle scattering if:
\begin{align}
	\ve{u} \cdot \ve{D} \cdot \pa{f}{\ve{u}} = 0 \quad \forall f.
\end{align}
Now, note that:
\begin{align}
	\ve{u} \cdot \ve{h}  = \bvec u_\perp\\ u_\parallel \evec \cdot \bvec u_\parallel \\ -u_\perp \evec  = \ve{0}.
\end{align}
Thus, for the $\ve{H}$ part of the $\ve{D}$ matrix, we have:
\begin{align}
	\ve{u} \cdot \ve{H} \cdot \pa{f}{\ve{u}} &= \left[\ve{u} \cdot \ve{h} \right] \left[ \ve{h} \cdot \pa{f}{\ve{u}} \right] = \ve{0}.
\end{align}
Thus, the $\ve{H}$ part of $\ve{D}$ represents pitch angle scattering solely.

As a corollary, any distribution $f$ that is a function only of the energy does not diffuse due to $\ve{H}$.
This can be seen by noting that any function of the energy is a function only of $\gamma = \sqrt{1 + u_\perp^2 + u_\parallel^2}$.
Thus, by the chain rule, the resultant diffusive flux is proportional to:
\begin{align}
	\ve{H} \cdot \pa{f}{\ve{u}} &= \ve{h} \ve{h} \cdot \frac{1}{\gamma} \bvec u_\perp \\ u_\parallel \evec \pa{f}{\gamma} = \ve{h} \left(\ve{h} \cdot \ve{u}\right) \frac{1}{\gamma}  \pa{f}{\gamma} = \ve{0}.
\end{align}

\subsection{Einstein Relation}

The Einstein relation says that when the electrons are in thermal equilibrium with the radiation ($T_\text{bb} = T_e$), the diffusive flux due to radiation absorption should balance the radiative drag.

In thermal equilibrium at temperature $T_e$, the electron distribution is given by the Maxwell-J\"uttner distribution:
\begin{align}
	f_\text{MJ} &\propto e^{-\gamma (mc^2/T_e)} .
\end{align}
Thus, 
\begin{align}
	\pa{f_\text{MJ}}{\ve{u}} &= \frac{\ve{u}}{\gamma} \pa{f}{\gamma} = -\ve{u} \frac{m c^2}{\gamma T_e} f_\text{MJ}.
\end{align}

Now we apply the diffusion operator to this distribution.
We already know that the $\ve{H}$ matrix does not contribute, since $f_\text{MJ}$ is only a function of the energy and not of the pitch angle.
Then, note that:
\begin{align}
	\ve{g} \cdot \ve{u} &= u_\perp^2 \left(1 + \frac{u_\parallel^2}{\gamma_\perp^2}  \right) = \frac{u_\perp^2 \gamma^2}{\gamma_\perp^2}
\end{align}
Thus, the diffusive flux is given by:
\begin{align}
	\ve{D}_\text{bb} \cdot \pa{f_\text{MJ}}{\ve{u}} &= \nu_R \frac{T_\text{bb}}{m c^2} \frac{\gamma_\perp^2}{\gamma} \frac{1}{u_\perp^2} \ve{G} \cdot \pa{f}{\ve{u}}\\
	&= - \nu_R \frac{T_\text{bb}}{m c^2} \frac{\gamma_\perp^2}{\gamma} \frac{1}{u_\perp^2} \ve{g} \lp \ve{g} \cdot \ve{u} \rp \frac{m c^2}{\gamma T} f_\text{MJ}\\
	&= - \nu_R \frac{T_\text{bb}}{T_e} \frac{\gamma_\perp^2}{\gamma} \frac{1}{u_\perp^2} \ve{g} f_\text{MJ}
\end{align}
When the radiation and plasma are in thermal equilibrium, $T_\text{bb} = T_e$, and comparison to Eq.~(\ref{eq:GammaBornatici}) yields:
\begin{align}
	\ve{D}_\text{bb} \cdot \pa{f_\text{MJ}}{\ve{u}} &= \Gammav f_\text{MJ}. \label{eq:EinsteinRelation}
\end{align}
Comparison with Eq.~(\ref{eq:FundamentalDiffusion}) shows that this implies that the diffusive flux vanishes.
Thus, the Einstein relation is satisfied for the diffusion operator in Eq.~(\ref{eq:DiffusionMatrixDecomposition}), regardless of the value of $\Delta$.

\subsection{Pitch Angle Diffusion Coefficient $\Delta$} \label{sec:Delta}

Calculating $\Delta$ requires the calculation of $U_2$, which was not given by Bornatici or Melrose.
However, we can follow the path laid out by Melrose to complete the calculation, which is reported in Appendix~\ref{app:DeltaCalc}.
The result is:
\begin{align}
	\Delta &= \frac{u_\perp^2}{16 \gamma_\perp^4 u^7} \times \notag\\
	& \hspace{10pt}  
	\biggl\{u\bigl[10 \gamma^4 \gamma_\perp^2 -38 \gamma^4 + 21 \gamma^2 \gamma_\perp^2 + 23 \gamma^2 - 16 \gamma_\perp^2 ] \notag\\
	& \hspace{12pt} + 3\gamma \arccosh (\gamma) \bigl[8\gamma^4 - 12\gamma^2 \gamma_\perp^2 + 7 \gamma_\perp^2 -3 \bigr]\biggr\}. \label{eq:DeltaFull}
\end{align}
The asymptotic limits are:
\begin{align}
	\frac{\Delta}{u_\perp^2} \rightarrow \begin{cases}
		\frac{2}{5} \quad & u_\parallel \sim u_\perp \rightarrow 0\\
		\frac{3}{2} \frac{\log(\gamma)}{\gamma^2 \gamma_\perp^4} \quad & u_\parallel\rightarrow \infty,u_\perp \lesssim 1.\\
		\frac{5}{8} \frac{1}{\gamma^2 \gamma_\perp^2} \quad & u_\parallel \sim u_\perp \rightarrow \infty
	\end{cases}
\end{align}

\begin{figure}
	\centering
	\includegraphics[width=\linewidth]{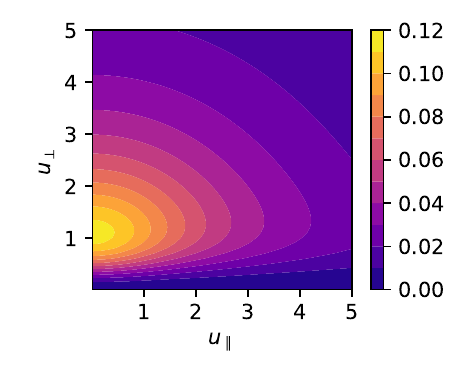}
	\caption{Pitch angle scatting coefficient $\Delta$ for blackbody synchrotron radiation as a function of $u_\parallel$ and $u_\perp$. 
	The pitch angle scattering is always less than 12\% as strong as the diffusion along the path of radiative drag.
	The pitch angle scattering has the largest relative effect for marginal values of $u_\perp \sim 1$, when the electron is relativistic enough to radiate at many harmonics, but not so relativistic that the headlight effect narrows the cone of radiation to the pitch angle.}
	\label{fig:Delta}
\end{figure}

Fig.~\ref{fig:Delta} shows $\Delta$ across a broad range of $u_\parallel$ and $u_\perp$. 
It can be seen that $\Delta$ reaches a maximum value around 12\% at marginal values of $u_\perp \sim 1$, falling to 0 in both the nonrelativistic and ultrarelativistic limits. 
This behavior can be understood as follows.
The pitch angle scattering results from the fact that diffusion occurs along a different paths $\ve{d^{(s)}}$ for each harmonic $s$.
In the nonrelativistic limit, almost all radiation is at the fundamental harmonic, so that there effectively only a single diffusion path.
In the ultrarelativistic limit, meanwhile, the radiation is emitted in a cone of order $|\theta-\alpha| \sim \gamma^{-1}$, so that the differences in the diffusion paths disappear, and the pitch angle scattering goes away.
Thus, as expected, pitch angle scattering only occurs for marginal values of $u_\perp,u \sim \mathcal{O}(1)$.

If one does not like transcribing long formulas, one can use the asymptotically-valid fit:
\begin{align}
	\Delta \approx  u_\perp^2\frac{60\log(\gamma) + 25 \gamma_\perp^2 -9}{16 \gamma^2 \gamma_\perp^4 + (a\gamma^b \gamma_\perp^c - a)}.
\end{align}
For $a = 49.81$, $b = 1.426$, $c = 3.568$, this fits to within 10\% for all values of $u_\parallel$ and $u_\perp$.

\section{Effective Diffusion Model for Mixed-Opacity Plasmas} \label{sec:DiffusionMixedOpacity}

To get a full sense of the radiative dynamics, one should properly solve the Fokker-Planck equation in conjunction with the global radiative transport equation.
However, it turns out that the scale length of self-absorption of synchrotron radiation is extremely frequency dependent, with low frequencies almost completely reabsorbed, and high frequencies almost completely escaping from the plasma.
Thus, it is a common approximation\cite{Trubnikov1958PlasmaRadiation,Drummond1963CyclotronRadiation,Trubnikov1961AngularDistribution,BekefiRadiationProcesses} to assume that the distribution is blackbody up to some cutoff frequency $\omega^*$, after which it quickly plunges to 0.

Even with this simplification, the calculation of the diffusion defies clean analytics.
In theory, one could carry out the calculation as in the blackbody case, but with the upper integration limit on $\omega$ in Eq.~(\ref{eq:DiffusionBornaticiBB}) changed to $\omega^*$.
However, this would not be very tractable, as the change in the limits of integration would ultimately break the Bessel sum identities that made the problem tractable in the first place.

Instead, we note that the calculation would be much simpler if, in Eq.~(\ref{eq:DiffusionBornaticiBB}), $\ve{d^{(s)}}$ was the same for every $s,\omega,\theta$.
Then we could write the moderate-opacity diffusion operator as $\ve{D}_\text{mo}$ as:
\begin{align}
	\ve{D}_\text{mo} &= \ve{D}_\text{bb} (1-\lambda(\ve{u})), \label{eq:DmoFromDbb}
\end{align}
where
\begin{align}
	\lambda(\ve{u}) \equiv \frac{\sum_{i,s} 2\pi  d\omega \int_{-1}^{1} d \cos \theta \int_{\omega^*}^\infty \eta_s^{(i)}}{\sum_{i,s} 2\pi  \int_{-1}^{1} d \cos \theta \int_0^\infty d\omega \eta_s^{(i)}} \label{eq:LambdaGeneral}
\end{align}
is the fraction of emissivity above the cutoff frequency $\omega^*$, i.e. the fraction of the emissivity that escapes the plasma.
Note that the integration order has $\omega$ after $\theta$, since in general $\omega^*$ can be a function of $\theta$.

As noted in Sec.~\ref{sec:Delta}, taking $\ve{d^{(s)}}$ to independent of $s,\omega,\theta$ is a good approximation in the nonrelativistic and ultrarelativistic limits.
From the $\delta$ function in $\eta^{(i)}$ (Eq.~\ref{eq:DeltaFunction}) and the definition of $\ve{d^{(s)}}$, we have:
\begin{align}
	\ve{d^{(s)}} &= \frac{1}{u_\perp} \lp \gamma - u_\parallel \cos \theta, u_\perp \cos \theta \rp
\end{align}
In the nonrelativistic limit:
\begin{align}
	\ve{d^{(s)}} \rightarrow \lp u_\perp^{-1}, 0 \rp,
\end{align}
which is trivially independent of $s,\omega,\theta$.
In the relativistic limit, we have
\begin{align}
	\ve{d^{(s)}} \rightarrow \frac{1}{u_\perp} \lp u - u_\parallel \cos \theta, u_\perp \cos \theta \rp,
\end{align}
which looks very dependent on $\theta$. 
However, in this limit, the headlight effect \cite{Melrose1980PlasmaAstrophysics} ensures that nearly all radiation is emitted at the particle pitch angle, so that $\cos \theta \approx u_\parallel / u$.
Plugging this in, we find:
\begin{align}
	\ve{d^{(s)}} \rightarrow \frac{1}{u} \lp u_\perp, u_\parallel \rp,
\end{align}
which is again independent of $s$ and $\omega$.
Thus, the model should work well except in detailed quantitative modeling of the pitch-angle diffusion at moderate $\lambda \sim 1/2$, when the frequency-dependence of $\Delta$ will be lost, and thus the (relatively weak) pitch angle scattering might be off by $\mathcal{O}(1)$.

It is easy to verify that the mixed-opacity diffusion coefficient behaves well with respect to the power balance.
Consider the net power emitted by the plasma in thermal equilibrium.
The net power radiated, from both emission and self absorption, can be calculated from Eq.~(\ref{eq:FundamentalDiffusion}) in units of $mc^2$ as:
\begin{align}
	P_\text{net} &= -\int \gamma \pa{f}{t} \sqrt{g} \\
	&= \int \frac{1}{\gamma} \bvec u_\perp \\ u_\parallel \evec  \cdot  \left(- \ve{\Gamma} f + \ve{D} \cdot \pa{f}{\ve{u}} \right)  \sqrt{g} d\ve{u}
\end{align}
Plugging in $\ve{D}_\text{mo}$ for $\ve{D}$ and using Eq.~(\ref{eq:EinsteinRelation}) for $\ve{D}_\text{bb}$:
\begin{align}
	P_\text{net} 
	&=  \int \frac{1}{\gamma} \bvec u_\perp \\ u_\parallel \evec  \cdot  \left(- \lambda \ve{\Gamma} f_\text{MJ} \right)  \sqrt{g} d\ve{u}.
\end{align}
Interpreting the last line, we recognize the expression for the total emitted power, modulated by the function $\lambda(\ve{u},\omega^*)$.
Thus, self-consistently, $\lambda(\ve{u},\omega^*)$ represents the fraction of emitted power that is not reabsorbed by the plasma.
In other words, despite the simplifications of the model, $\ve{D}_\text{mo}$ sensibly respects energy conservation.

To complete the model, we must now calculate $\lambda(\ve{u},\omega^*)$.
We must also calculate $\omega^*$ itself, which depends on the self-absorption of the plasma.
We call these functions the ``opacity functions'' of the model.

\section{Opacity Functions for a Mildly Relativistic Plasma} \label{sec:OpacityFunctions}

\subsection{Cutoff Frequency $\omega^*$}

The typical model to calculate the frequency cutoff for absorption is to consider a slab (or cylinder) of plasma at a uniform temperature.
Starting from the center, photons will be absorbed on a scale length $1/\alpha(\omega)$, which typically depends very strongly on $\omega$.
Thus, photons with frequencies above $\omega^*$ largely escape the plasma, while photons with frequencies below $\omega^*$ are largely trapped, with $\omega^*$ given by solving $\alpha(\omega^*) L = 1$.
This model neglects some subtleties associated with the angle at which the photon leaves the plasma, which should have negligible impact anyway.

Usually, for mildly relativistic plasmas, the cutoff is posed in terms of a cutoff harmonic $m^*$ rather than cutoff frequency $\omega^*$, such that there is no self-absorption for $s \geq m^*$.
Trubnikov and Kudryatsev\cite{Trubnikov1958PlasmaRadiation} proposed a simple analytic model, which was largely verified by the more intensive calculations of Drummond and Rosenbluth.\cite{Drummond1963CyclotronRadiation}
This model gives $m^*$ as the solution to:
\begin{align}
	1 = \Lambda \frac{3 \pi^{1/2}}{2 \kappa \chi_e^{1/2}} \exp \left[\chi_e^{-1} \lp \kappa^{1/3} - 1 + \tfrac{9}{20} \kappa^{-1/3} \rp \right], \label{eq:MStarTrubnikov}
\end{align}
where
\begin{align}
	\Lambda &\equiv L \frac{\omega_{pe}^2}{c \Omega} = \frac{L}{L_\text{sd}} \frac{\omega_{pe}}{\Omega} ; \;
	\chi_e \equiv \frac{T_e}{m_e c^2}; \; 
	\kappa \equiv \frac{9}{2} \chi_e m^*.
\end{align}
Here, $T_e$ is the electron temperature, $\omega_{pe}$ is the electron plasma frequency, and $L_\text{sd}$ is the plasma skin depth.
The model is valid in the limit $\chi_e \equiv T_e/m_e c^2 \ll 1$ and $\kappa \gtrsim 1$.

\begin{figure}
	\centering
	\includegraphics[width=\linewidth]{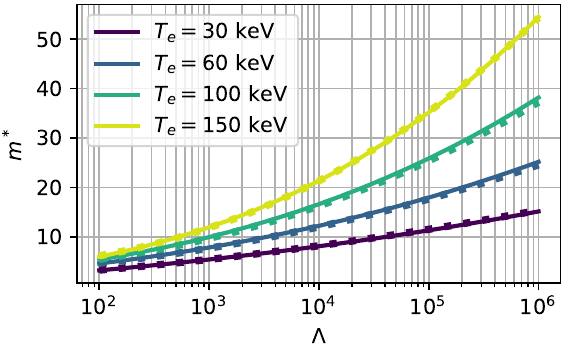}
	\caption{Cutoff harmonic $m^*$ as a function of $\Lambda$ for several values of $T_e$ as calculated from the numerical solution to the Trubnikov-Kudryatsev model in Eq.~(\ref{eq:MStarTrubnikov}) [solid], versus from the analytic approximation in Eqs.(\ref{eq:MStarFit}-\ref{eq:MStarParamV2}) [dashed].}
	\label{fig:MildRelativisticMStar}
\end{figure}

A consequence of Eq.~(\ref{eq:MStarTrubnikov}) is that $m^*$ is a function only of $\Lambda$ and $\chi_e$, which can be easily numerically solved.
Fig.~\ref{fig:MildRelativisticMStar} shows $m^*$ as a function of $\Lambda$ for several values of $T_e$.
While Ref.~\onlinecite{BekefiRadiationProcesses} offers an approximate fit for the special case of $T_e = 50$ keV, modern computers allow for a much more accurate 9-parameter fit across a broad range of (30~keV~$< T_e <$~200~keV) and ($10^2 < \Lambda < 10^6$).
This fit is given by:
\begin{align}
	m^* &= v_0(T_e) + v_1(T_e) \log(\Lambda)^{v_2(T_e)}. \label{eq:MStarFit}
\end{align}
The parameter functions are fit by:
\begin{align}
	v_0(T_e) &= w_0 \log(T_e w_1 + w_2)\\
	v_1(T_e) &= w_3/T_e^{w_4}\\
	v_2(T_e) &= w_5 \log(T_e w_6 + w_7), \label{eq:MStarParamV2}
\end{align}
where $T_e$ is given in keV.
Figure~\ref{fig:MildRelativisticMStar} shows the model alongside the numerical solution of Eq.~(\ref{eq:MStarTrubnikov}) for several values of $T_e$, where the fit is seen to be quite accurate.

\begin{table}[b]
	\centering
	\begin{tabular}{| c | c | c |}
		\hline
		\; $w_{0}$ = 5.75e-01 \; & \; $w_{1}$ = 1.49e+00 \; & \; $w_{2}$ = -3.75e+01 \; \\ 
		\hline 
		\; $w_{3}$ = 8.99e-01 \; & \; $w_{4}$ = 5.26e-01 \; & \;  \; \\ 
		\hline 
		\; $w_{5}$ = 6.71e-01 \; & \; $w_{6}$ = 2.59e-01 \; & \; $w_{7}$ = 5.46e+00 \; \\ 
		\hline 
	\end{tabular}
	\caption{Parameters for analytical approximation [Eqs.~(\ref{eq:MStarFit}-\ref{eq:MStarParamV2})] to the solution of Eq.~(\ref{eq:MStarTrubnikov}) for $m^*$.}
	\label{tab:MStarFitParametersTK}
\end{table}

\subsection{Radiation Escape Fraction $\lambda$}

With the cutoff frequency $m^*$ established, we now turn to calculating the radiation escape fraction $\lambda(\ve{u})$.
As discussed in Appendix~\ref{app:DeltaCalc}, the power radiated from a section of phase space at $\ve{u}$ is proportional to
\begin{alignat}{1}
	I_{0} &\equiv  \sum_{s=1}^{\infty} \int_{0}^{1} d w  \sum_{i=1}^2  s^2 F^{ii}, \label{eq:EmmissionSumS}
\end{alignat}
where
\begin{align}
	F^{11} &= \left(\frac{w}{\beta' \sqrt{1-w^2}}\right)^2 \left[J_s\left(s \beta' \sqrt{1-w^2}\right)\right]^2;\\
	F^{22} &=  \left[J_s'\left(s \beta' \sqrt{1-w^2}\right)\right]^2
\end{align}
are the emission associated with the ordinary (O) and extraordinary (X) modes respectively,
and $\beta' = u_\perp / \gamma_\perp$.
Properly, there is a different effective cutoff $m^*$ associated with each of these modes; however, following prior authors,\cite{BekefiRadiationProcesses} we will brush over this distinction, which does not make a large difference to the final result and reduces the computational complexity significantly.

Examining Eq.~(\ref{eq:EmmissionSumS}), it is clear that the fraction of power lost is given by:
\begin{align}
	\lambda = \frac{\sum_{s=m^*}^{\infty} \int_{0}^{1} d w  \sum_{i=1}^2  s^2 F^{ii}}{\sum_{s=1}^{\infty} \int_{0}^{1} d w  \sum_{i=1}^2  s^2 F^{ii}} . \label{eq:LambdaFullCalc}
\end{align}
This quantity can be calculated numerically straightforwardly; however, it is useful to derive a fit, especially because as the plasma grows more relativistic, hundreds of harmonics are required to calculate the result.

\begin{table}[b]
	\centering
	\begin{tabular}{| c | c | c |}
		\hline
		\; $y_{0}$ = 3.11e-01 \; & \; $y_{1}$ = 4.80e+00 \; & \; $y_{2}$ = 3.49e+00 \; \\ 
		\hline 
		\; $y_{3}$ = -9.08e-04 \; & \; $y_{4}$ = 3.58e+00 \; & \; $y_{5}$ = -1.72e+01 \; \\ 
		\hline 
		\; $y_{6}$ = 8.00e-03 \; & \; $y_{7}$ = 2.28e-01 \; &  \\ 
		\hline 
		\; $y_{8}$ = 7.14e-02 \; & \; $y_{9}$ = 2.55e+00 \; & \; $y_{10}$ = -3.53e+00 \;  \\ 
		\hline 
		\; $y_{11}$ = 6.18e+00 \; & \; $y_{12}$ = 6.12e-01 \;  &\\
		\hline
	\end{tabular}
	\caption{Parameters for analytical approximation to $\lambda$ [Eqs.(\ref{eq:LambdaMildRelApprox}-\ref{eq:LambdaMildRelParamX2})].}
	\label{tab:LambdaFitParametersMildRelativistic}
\end{table}

For any given value of $m^*$, $\lambda$ as a function of $u_\perp$ resembles a standard asymmetric logistic curve, known as the 5-parameter logistic curve, which is common in the study of dose response in biology.\cite{Gottschalk2005FiveparameterLogistic}
Since $\lambda \rightarrow 0$ as $u_\perp \rightarrow 0$, and $\lambda \rightarrow 1$ as $u_\perp \rightarrow \infty$, the 5-parameter curve becomes a 3-parameter curve, given by:
\begin{align}
	\lambda_\text{fit} &= \left( 1 + \left(\frac{u_\perp}{x_0(m^*)}\right)^{-x_1(m^*)} \right)^{-x_2(m^*)}. \label{eq:LambdaMildRelApprox}
\end{align}
The parameter functions $x_i(m^*)$ can then be fit, by:
\begin{align}
	x_0(m^*) &= y_0 \log(y_1 m^*+y_2)\\
	x_1(m^*) &= y_3 m^* + y_4 + y_5\exp\left(-(m^*/y_6)^{y_7}\right)\\
	x_2(m^*) &= y_8 m^* + y_9 + y_{10}\exp\left(-(m^*/y_{11})^{y_{12}}\right). \label{eq:LambdaMildRelParamX2}
\end{align}

\begin{figure}
	\centering
	\includegraphics[width=\linewidth]{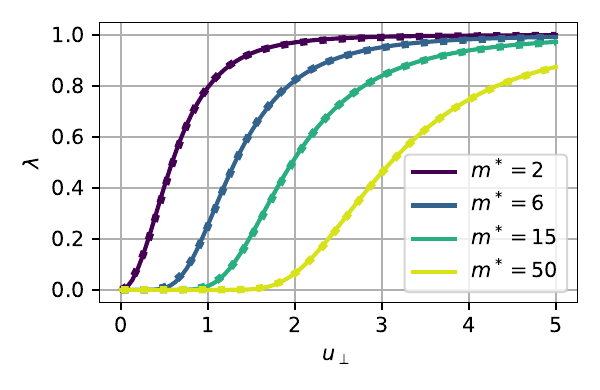}
	\caption{Radiation escape fraction $\lambda$ as a function of normalized perpendicular momentum $u_\perp$ for several values of the cutoff harmonic $m^*$, as calculated from the full Bessel function product sum in Eq.~(\ref{eq:LambdaGeneral}) over 1000 harmonics [solid] versus from the analytical approximation in Eq.~(\ref{eq:LambdaMildRelApprox}-\ref{eq:LambdaMildRelParamX2}) [dashed].}
	\label{fig:MildRelativisticLambda}
\end{figure}

Figure~\ref{fig:MildRelativisticLambda} shows the fit for $\lambda(u_\perp)$ alongside the numerical value from Eq.~(\ref{eq:LambdaFullCalc}) for several values of $m^*$; the agreement is seen to be quite good.
However, it is important to note that this is a good fit only when there are significant losses ($\lambda \gtrsim 0.02$), corresponding to the case when the primary losses are from the relativistic tails of the distribution function.
Thus, the model fit is not good for evaluating total radiative losses in the optically thick regime, when it tends to overestimate the (negligible) losses substantially.
Of course, in the optically-thick limit, an optically-mixed Fokker-Planck model is not necessarily the correct tool.
Nevertheless, in case this behavior is required, we note that in the limit $u_\perp \ll 1$, $\lambda$ asymptotically approaches [see App.~\ref{app:NonrelativisticLambda}]:
\begin{align}
	\lambda \rightarrow \lambda_\text{asym} \equiv m^{1/2} \left(\frac{e}{2} \frac{u_\perp}{\gamma_\perp}\rp^{2(m-1)}.
\end{align}

\section{Summary of Mixed-Opacity Synchrotron Model} \label{sec:Summary}

It is worthwhile here to summarize the components of the mixed-opacity synchrotron absorption-emission model.
The emission and absorption of radiation are modeled as drag and diffusion terms in a Fokker-Planck equation [Eq.~(\ref{eq:FundamentalDiffusion})].
The well-known drag $\ve{\Gamma}$ is given by Eq.~(\ref{eq:GammaBornatici}).
The diffusion is given from Eq.~(\ref{eq:DmoFromDbb}) as a function of the newly-solved-for blackbody diffusion operator in Eq.~(\ref{eq:DiffusionMatrixDecomposition}), with the new pitch-angle-scattering coefficient $\Delta$ given by Eq.~(\ref{eq:DeltaFull}). 
Thus, we have:
\begin{align}
	\Gammav &= -\nu_{R0} \frac{\gamma_\perp^2}{\gamma} \ve{g} \label{eq:GammaXTheta}\\
	\ve{D}_\text{mo} &= (1-\lambda(\ve{u})) \nu_{R0} \chi_\text{bb} \left( \frac{\gamma_\perp^2}{\gamma} \right)^2 \frac{1}{u_\perp^2} \left(\ve{g}\ve{g} + \Delta \ve{h}\ve{h} \right), \label{eq:DXTheta}
\end{align}
where $\chi_\text{bb} = T_\text{bb} / m_e c^2$ and $\nu_{R0}$ is given in Eq.~(\ref{eq:nuR0}).

The model relies on a phase-space dependent opacity function $\lambda(\ve{u},m^*)$, representing the fraction of emitted radiation that escapes the plasma, which is approximately fit for mildly relativistic plasmas by Eqs.~(\ref{eq:LambdaMildRelApprox}-\ref{eq:LambdaMildRelParamX2}).
The function $\lambda$ in turn depends on the cutoff harmonic $m^*$, which depends on the physical dimensions and plasma parameters of the surrounding medium.
The cutoff harmonic $m^*$ is given approximately as the solution to Eq.~(\ref{eq:MStarTrubnikov}), which can be approximated by Eqs.~(\ref{eq:MStarFit}-\ref{eq:MStarParamV2}).

Having defined the model, we now proceed in Secs.~\ref{sec:QuasiSteadyState}-\ref{sec:MirrorConfinement} to study the effects of synchrotron emission and absorption on mildly relativistic plasmas.

\section{Simulation Coordinates and the Quasi-Steady State} \label{sec:QuasiSteadyState}

\begin{figure*}[t]
	\centering
	\includegraphics[width=\linewidth]{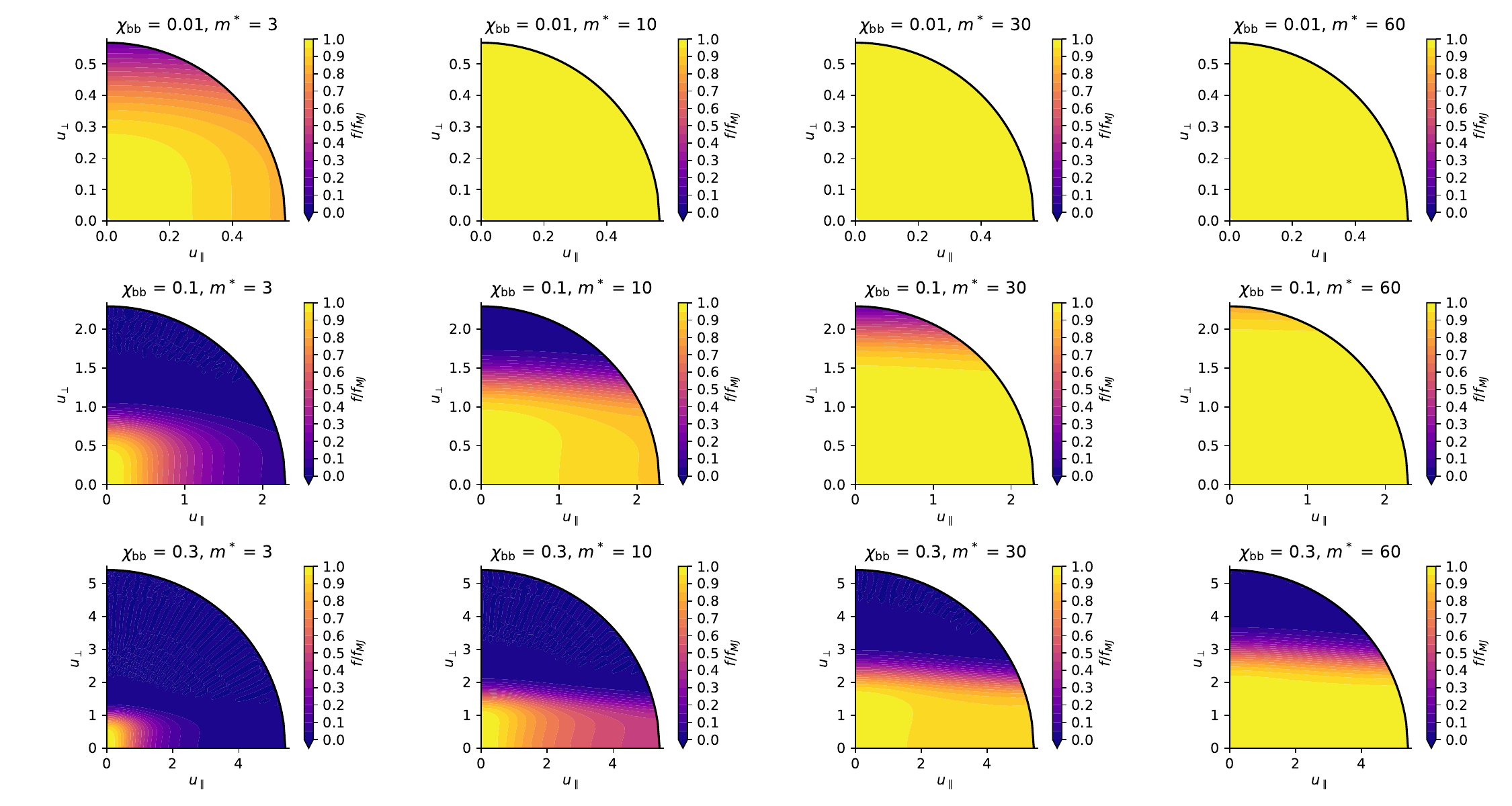}
	\caption{Quasi-steady state solution to the synchrotron Fokker-Planck operator, for several values of radiation temperature $\chi_\text{bb} = T_\text{bb} / m_e c^2$ and cutoff harmonic $m^*$.
	The solution is plotted as a ratio between the actual value of the distribution relative to the Maxwell-J\:uttner distribution, which is the steady-state solution in the optically thick limit.
	Plot limits extend to 15 times the blackbody temperature.
	For nonrelativistic temperatures $\chi_\text{bb} \ll 1$, where most radiation is emitted and absorbed at low harmonics, the plasma is optically thick even for very low values of the cutoff harmonic, and quickly approaches a Maxwell-J\"uttner distribution with increasing $m^*$.
	However, as the plasma grows more relativistic, more radiation escapes, and the distribution is suppressed, primarily in $u_\perp$ but also (to a lesser extent) in $u_\parallel$.
	Note: for optically thin plasmas, the drag becomes much stronger than the diffusion, and artifacts of numerical instability begin to appear at large $u_\perp$.}
	\label{fig:ClosedSystemSims}
\end{figure*}

To study the effect of synchrotron absorption and emission, we will study the steady-state in a system with a fixed blackbody temperature $T_\text{bb}$ and cutoff harmonic $m^*$, using the DOLFINx-\cite{Scroggs2022ConstructionArbitrary} and Gmsh-based\cite{Geuzaine2009Gmsh3D} Fokker-Planck framework from Ref.~\onlinecite{Ochs2023ConfinementTime}.
we first translate the diffusion and drag operators to coordinates $(x,\theta)$, where 
\begin{align}
	u_\perp = \sqrt{2 \chi} x \sin \theta; \quad u_\parallel = \sqrt{2 \chi} x \cos \theta.
\end{align}
Here, $\chi = T_0 / m_e c^2$ is the temperature that sets the normalization of the grid, usually taken as the temperature of the background electrons with which collisions take place.
As outlined in Appendix~\ref{app:CoordinateConversion}, the diffusion and drag terms take the same form as in Eqs.~(\ref{eq:GammaXTheta}-\ref{eq:DXTheta}), as long as $\ve{g}$ and $\ve{h}$ are redefined as:
\begin{align}
	\ve{g} \rightarrow \gamma_\perp^{-2} \bvec \gamma^2 x \sin^2 \theta \\ \cos\theta \sin \theta \evec ; \quad \ve{h} \rightarrow \bvec 0 \\ 1 \evec.
\end{align}
[Note that for the choice $\chi = 1/\sqrt{2}$, $x = u$, and so the above operator works in $(u,\theta)$ space as well.]

Using this Fokker-Planck operator, we can solve for the steady state of a radiating system, where the value of $f(x,\theta)$ near the $x = 0$ boundary is fixed, while the flux is set to 0 at all other boundaries.
Now, this is not a true steady state of the radiating plasma, since of course the plasma becomes more transparent as it cools, eventually losing all of its energy; however, it represents a quasi-steady state of the particle distribution given fixed radiation properties of the plasma, and thus reveals the tendencies of the mixed-opacity synchrotron operator.
The value of $f$ for this ``quasi-steady'' state can then be compared to a Maxwell-J\"uttner distribution.

Figure~\ref{fig:ClosedSystemSims} shows the resulting simulations, for several values of the blackbody temperature $\chi_\text{bb} \equiv T_\text{bb} / m_e c^2$ and several values of the cutoff harmonic $m^*$.
It is clear that as the cutoff harmonic increases and the plasma grows more opaque, more of the phase space becomes close to the Maxwell-J\"uttner distribution.
However, it is also clear that radically more harmonics are required as the perpendicular momentum of the plasma grows more relativistic. 
Indeed, for $u_\perp > 2$, even 60 harmonics are not sufficient to avoid significant suppression of the tail relative to a Maxwell-J\"uttner distribution.
Combined with the relative weakness of the pitch-angle scattering, this $u_\perp$ suppression means that the tendency is for synchrotron radiation to produce asymmetric distributions that extend further into $u_\parallel$ than $u_\perp$, with $f$ heavily suppressed for $u_\perp \gtrsim 2$.

\section{Effect on Equilibrium and Confinement in Magnetic Mirrors} \label{sec:MirrorConfinement}

In a realistic laboratory fusion plasma, the synchrotron emission and absorption act alongside collisions to form a radiative-collisional equilibrium.
For aneutronic fusion plasmas, the synchrotron radiation can have a very strong effect on the resulting equilibrium, which can deviate strongly from a Maxwell-J\"uttner distribution.
As a result, the net power radiated from the plasma can be very different from the power calculated assuming local thermodynamic equilibrium.

The details of the radiative-collisional equilibrium are particularly important for the mirror confinement of aneutronic fusion plasmas, since both the radiative losses and the electron confinement properties are determined by the high-energy tails of the electron distribution.


The relativistic collisions for tail electrons can be numerically modeled by a Fokker-Planck operator with drag vector and diffusion tensor given by:\cite{Ochs2023ConfinementTime} 
\begin{align}
	\Gammav_\text{coll} &= -\frac{\gamma^2 x}{\tau_0 (x')^3} \hat{x} \label{eq:GammaCollXTheta}\\
	\Dv_\text{coll} &= \frac{\gamma^3}{2 \tau_0 (x')^3} \hat{x}\hat{x} + \frac{\gamma }{\tau_0 (x')^3} \lp Z_\perp - \frac{Z_\perp}{1 + c + 4 Z_\perp x^2} \rp \hat{\theta} \hat{\theta}
\end{align}
where
\begin{align}
	\tau_0^{-1} &\equiv 4\pi e^4 \frac{n_e \lambda_{ee} m_e}{p_{\text{th},e}^3},\\
	Z_\perp &= \frac{1}{2} \frac{\sum_b n_b Z_b^2 \lambda_{eb} }{\sum_b n_b Z_b^2 \lambda_{eb} m_e T_b / m_b T_e},
\end{align}
and $(x')^3 = x^3 + x_0^3$, and where here we take $x_0 = 0.1$ and $c = 0.2$.

Usually, the simulations are time-normalized to the electron collision time.
Thus, we take $\tau_0 \rightarrow 1$, and in the synchrotron operator we take $\nu_{R0} \rightarrow \bar{\nu}_{R0}$, where:
\begin{align}
	\bar{\nu}_{R0} \equiv \nu_{R0} \tau_0 = \frac{2}{3} \frac{\Omega_0^2}{\omega_{pe}^2} u_\text{th}^3 \lambda_{ee}^{-1},
\end{align}
with $u_\text{th} \equiv \sqrt{2 m_e T_e}/m_e c$.
For a thermonuclear p-B11 plasma, with 300 keV ions and 150 keV electrons, at a total ion density of $10^{14}$ cm$^{-3}$ and a magnetic field of 10T, $\bar{\nu}_{R0} \approx 0.1$.

The condition for the synchrotron drag to become comparable to the collision drag comes from dividing Eq.~(\ref{eq:GammaXTheta}) by Eq.~(\ref{eq:GammaCollXTheta}), which yields the parameter from Eq.~(\ref{eq:BornaticiParameter}):
\begin{align}
	\epsilon_S \equiv \frac{2}{3} \frac{\Omega_0^2}{\omega_{pe}^2} \frac{u^3}{\lambda_{ee} \gamma} = \bar{\nu}_{R0} \frac{x^3}{\gamma} \gtrsim 1.
\end{align}
Thus, for the pB11 fusion plasma above, we expect significant synchrotron effects at $x = 10^{1/3} \approx 2$ times the thermal momentum, i.e. not very far out on the tail. 

\begin{figure*}[t]
	\centering
	\includegraphics[width=\linewidth]{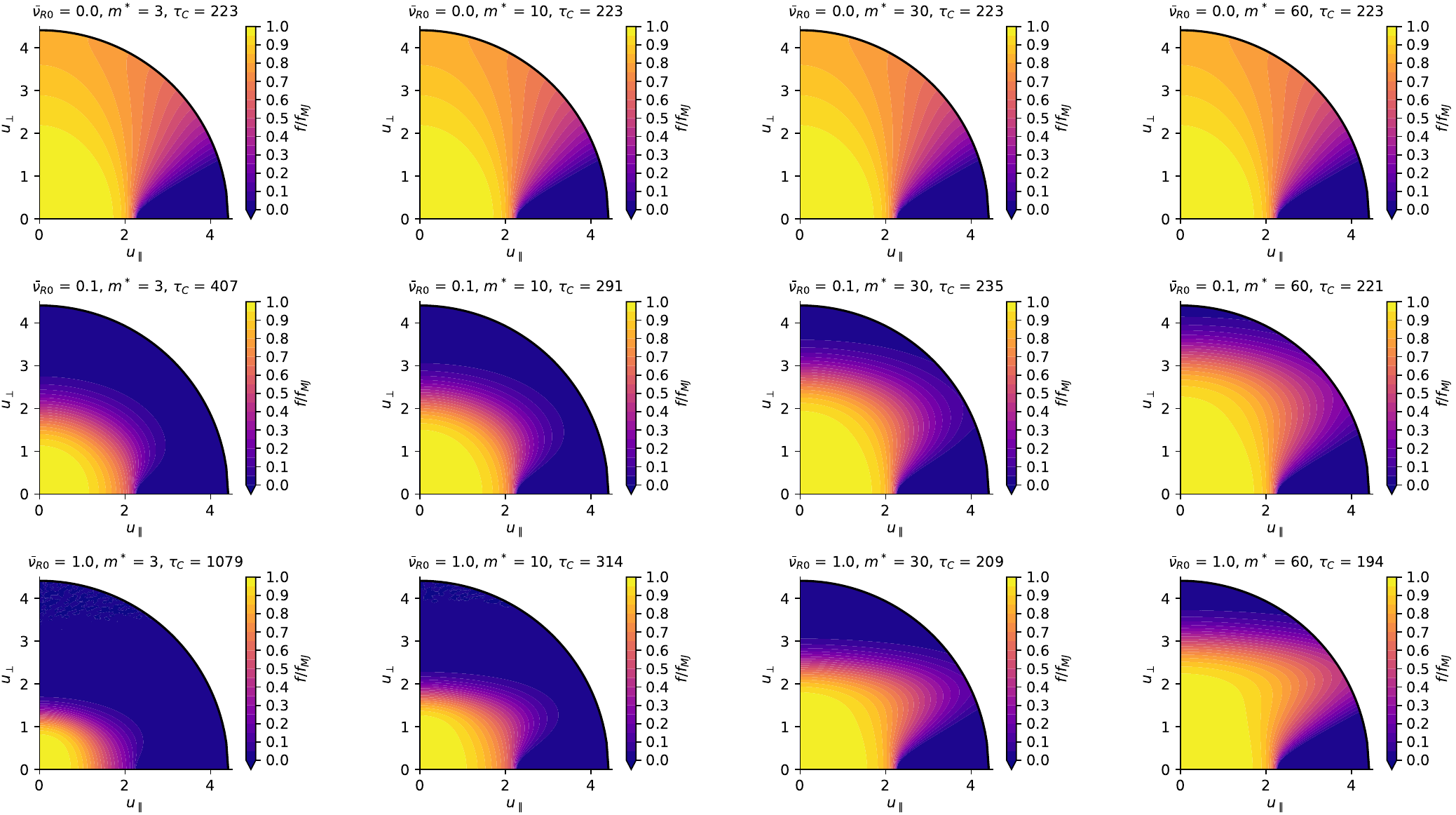}
	\caption{Steady-state electron distributions (relative to a Maxwell-J\"uttner distribution) in an aneutronic fusion magnetic mirror fusion scenario for several values of the synchrotron strength $\bar{\nu}_{R0}$ and cutoff harmonic $m^*$.
	The mirror is characterized by a mirror ratio $R = 5$ and ambipolar potential $\phi = 5 T_e$,
	Modeled are (a) a cold electron source, (b) relativistic collisions with $T_e = 150$ keV electrons and a mix of 300 keV protons (85\%) and Boron (15\%), and (c) synchrotron emission-absorption given a blackbody radiation temperature of $T_\text{bb} = T_e = 150$ keV.
	The resulting confinement times $\tau_C \equiv n_e / (dn_e/dt)$ (normalized to the electron collision time) are listed above each solution.
	The radiation has two opposite effects on the confinement: (1) it suppresses the tails through non-reabsorbed radiation drag, increasing the confinement time, and (2) it increases the effective electron collision frequency, lowering the confinement time.
	The tail suppression effect wins out in plasmas with sufficiently optically thin electron tails, while the collision effect wins out in optically thick plasmas.}
	\label{fig:MirrorSims}
\end{figure*}

Simulations are shown in Fig.~\ref{fig:MirrorSims} for electrons in magnetic mirror with a confining potential of $\phi = 5 T_e$ and a mirror ratio of $R = 5$, for a pB11 thermonuclear fusion scenario, in the presence of a cold electron source ($\llangle x_s \rrangle \sim 0.1$), relativistic collisions, and synchrotron radiation.
The resulting loss hyperboloid of the mirror is visible in the bottom right of each figure.
Simulations were performed for several values of the cutoff harmonic $m^*$ and the normalized synchrotron timescale $\bar{\nu}_{R0}$.
As in Fig.~\ref{fig:ClosedSystemSims}, the electron distribution is scaled to the Maxwell-J\"uttner distribution.
It is clear the the synchrotron radiation significantly suppresses the tails of the distribution at high $u_\perp$.
For the most pB11-relevant case of $\bar{\nu}_{R0} = 0.1$, $m^* = 10$, the distribution is surprisingly fairly isotropic, with a rapid decline around $|u| = 2$, making the truncated Maxwell-J\"uttner model\cite{Mlodik2023SensitivitySynchrotron} potentially a fairly accurate way to estimate the synchrotron emission.

Also plotted in the title of each figure is the typical confinement time of electrons in the mirror (in units of the electron collision time $\tau_0$), which is defined as the integrated density divided by the integrated source rate ($\tau_C \equiv n_e / (dn_e/dt)$).
Interestingly, the confinement time can either increase or decrease as a result of the synchrotron emission and absorption.
However, this behavior can be understood very simply.
As noted in Refs.~\onlinecite{White2018CentrifugalParticle,Ochs2023ConfinementTime}, the collision time scales roughly as:
\begin{align}
	\tau_C \sim \frac{\tau_\perp}{f_L},
\end{align}
where $\tau_\perp$ is the perpendicular collision time, and $f_L$ is the fraction of the particle distribution function at an energy above the confining potential $\phi$ in thermal equilibrium.
The synchrotron emission and absorption reduces both the numerator, by enhancing the effective collisionality of the plasma, and the denominator, by suppressing the tails (where the plasma is more optically thin). 
Thus, at higher $m^*$, i.e. for more opaque plasmas, the synchrotron absorption can actually reduce the particle confinement time, by increasing the effective collisionality of the plasma.
However, in mildly relativistic plasmas of reactor-relevant scales, the tail suppression effect tends to win out, and thus we see an increase in the confinement time for the most reactor-relevant case of $\bar{\nu}_{R0} = 0.1$, $m^* = 10$.

\section{Discussion and Conclusion} \label{sec:Conclusion}

Here, we have developed a theory of electron kinetics in the presence synchrotron emission and absorption in mildly relativistic plasmas.
We started by completing the derivation of the blackbody diffusion operator, which was begun in Ref.~\onlinecite{Bornatici1994RelativisticFokkerPlanck}
We found that the diffusion operator could be understood as consisting of two terms: (i) a thermalizing component enforcing the Einstein relation along the radiative drag path, and (ii) a (novel) pitch-angle scattering component, primarily affecting particles at marginal values of $u_\perp \equiv p_\perp / mc \sim 1$.

With the blackbody diffusion coefficient in hand, we worked to find a minimal-complexity model to capture the diffusion in the regime of frequency-dependent plasma opacity that is characteristic of mildly relativistic laboratory-scale plasmas.
The requisite diffusion operator was constructed by scaling the blackbody diffusion tensor at each point in phase space by the fraction of emitted radiation $\lambda(\ve{u})$ that escapes from the plasma.
This diffusion operator sensibly respects energy conservation at each point in phase space, while being simple to calculate and implement.

Using the newly developed diffusion model, we examined the effect of the synchrotron operator on mildly relativistic electron plasmas, in conjunction with the relativistic Coulomb collision operator.
We found two main effects: (i) a suppression of the high-$u_\perp$ electron tail, and (ii) an increase in the effective collisionality of the plasma.
The tail suppression tends to reduce radiation from the plasma (relative to a Maxwell-J\"uttner distribution).
In a magnetic mirror configuration, the tail suppression tends to improve confinement, while the increased collisionality tends to degrade it; however, for typical parameters for a laboratory-scale aneutronic fusion plasma, the tail suppression effect wins out.

The distributions calculated here could be used in power balance calculations for aneutronic fusion, as an input into radiation loss functions, to get a more realistic calculation of the radiation losses in such plasmas.
However, the approximate isotropy of the distribution in the most fusion-relevant cases, at least for magnetic mirrors, suggests that a truncated-Maxwell-J\"uttner model could yield fairly accurate predictions.\cite{Mlodik2023SensitivitySynchrotron}

It should be noted that the current model assumes that the opacity and radiation properties of the plasma are fixed at a given temperature.
It would be interesting, in future work, to adapt the absorption properties of the plasma based on the kinetic state---however, this would make the system highly nonlinear, and thus much more computationally expensive.

Despite its simplifications, we note that the present model of synchrotron radiation also represents an improvement over the existing treatment of synchrotron radiation in quasilinear diffusion models of current drive using relativistic electrons, which tend to either ignore radiation completely \cite{Fisch1981CurrentGeneration} or treat the radiation emission but ignore the absorption.\cite{Harvey1992CQL3DFokkerplanck}
Though the existing models offer approximate upper and lower bounds on the current drive efficiency, they neglect the effect of radiation-induced diffusion.
While reasonable for current drive by nonrelativistic electrons, these effects are likely to become important for current drive by relativistic electrons even in DT fusion plasmas.
For instance, a 20 keV DT fusion plasma with a 10T magnetic field and 1m radius will trap harmonics below $m^* \sim 3$.
Thus, if the current is being driven in electrons of $u_\perp \sim 1$, it is both true that (a) synchrotron radiation is similarly important to collisions (with $\epsilon_S$ equal to 0.7) and (b) reabsorption is significant, with $\lambda = 0.6$.
Thus, diffusion terms in the Fokker-Planck equation could dramatically impact the current drive efficiency in this case, increasing the efficiency relative to the overly pessimistic emission-only case.
Of course, the radiation spectrum emitted by such electrons would be likely to be asymmetric along the direction of current drive, which would need to be incorporated into the FP model if the details of the thermal radiation reabsorption are important, as in synchrotron-reflection-based current drive.\cite{Dawson1982CurrentMaintenance}
Intriguingly, in a scenario in which synchrotron radiation were predominantly reabsorbed by high-energy electrons, the current drive efficiency could even be increased due to the radiation, via its transfer to less collisional electrons.

Finally, we note that radiative drag has recently been proposed as a driver of instabilities in ultra-relativistic laboratory and astrophysical plasmas.\cite{Zhdankin2023SynchrotronFirehose,Bilbao2023RadiationReaction,Bilbao2024RingMomentum}
The tendency observed in such systems is for the distribution to form a ring distribution at high $u_\perp$.
However, existing analyses have so far assumed an optically thin plasma.
The present paper provides a roadmap that could be used to extend the existing analysis to optically mixed or optically thick ultrarelativistic plasmas, via calculation of the ultrarelativistic opacity functions.

\section*{Acknowledgments}

This work was supported by ARPA-E Grant No. DE-AR0001554, and DOE Grant No. DE-SC0016072.

\appendix

\section{Calculation of Pitch-Angle Scattering Coefficient $\Delta$} \label{app:DeltaCalc}

We are interested in the quantities:
\begin{align}
	U_m = \sum_{i,s} 2\pi \int_0^\infty d\omega \int_{-1}^{1} d \cos \theta \left(\frac{s \Omega_0}{\omega}\right)^m\eta_s^{(i)}.
\end{align}
Now, Melrose writes his power as a polarization matrix, but it is only the diagonal components that represent the emitted power in each polarization (See discussion around Melrose Eq.~(3.19)).
Thus, in Melrose's notation, we are interested in the components:
\begin{align}
	U_m = \sum_{i=1}^2 \sum_{s=1}^{\infty} 2\pi \int_0^\infty d\omega \int_{-1}^{1} d \cos \theta \left(\frac{s \Omega_0}{\omega}\right)^m\eta^{ii}(s,\omega,\theta).
\end{align}

First, we can make use of the delta function relation (Eq.~\ref{eq:DeltaFunction}) to write:
\begin{align}
	\frac{s \Omega_0}{\omega} &= \gamma - u_\parallel \cos \theta = \gamma (1 - \beta_\parallel \cos \theta),
\end{align}
yielding
\begin{align}
	U_m = \sum_{i=1}^2 \sum_{s=1}^{\infty} 2\pi \int_0^\infty d\omega \int_{-1}^{1} d \cos \theta \gamma^m (1 - \beta_\parallel \cos \theta)^m\eta^{ii}(s,\omega,\theta).
\end{align}

Now, From Melrose~(4.39) and (4.40):
\begin{align}
	U_m = \frac{3}{2} C \beta_\perp^2 \gamma^{m-2} \sum_{i=1}^2 \sum_{s=1}^{\infty} \int_{-1}^{1} d \cos \theta  \frac{s^2 F^{ii}}{(1-\beta_\parallel \cos \theta)^{3-m}},
\end{align}
where $C = \nu_{R0} m c^2$ and
\begin{alignat}{3}
	F^{11} &= \left[\left(\frac{z}{x}\right) J_s(s x)\right]^2; \quad &F^{22} &= \left[ J'_s(s x)\right]^2;\\
	x &= \frac{\beta_\perp \sin \theta}{1 - \beta_\parallel \cos \theta}; \quad &z &= \frac{\cos \theta - \beta_\parallel}{1 - \beta_\parallel \cos \theta}. 
\end{alignat}

Continuing to follow Melrose [Eq.~(4.41)], switch variables to $\cos \theta'$, where:
\begin{align}
	\cos \theta' &= \frac{\cos \theta - \beta_\parallel}{1-\beta_\parallel \cos \theta}; \\
	\cos \theta &= \frac{\beta_\parallel + \cos \theta'}{1+\beta_\parallel \cos \theta'}\\
	d \cos \theta &= \frac{1 - \beta_\parallel^2}{(1 + \beta_\parallel \cos \theta')^2}.
\end{align}
Under $u$-substitution, the limits of integration remain unchanged, and so:
\begin{align}
	U_m &= 3 C \beta_\perp^2 (1-\beta_\parallel^2)^{m-2}\gamma^{m-2} I_m \label{eq:UmFromIm1}\\
	I_m &\equiv \frac{1}{2}\int_{-1}^{1} d \cos \theta'  (1+\beta_\parallel \cos \theta')^{1-m} \sum_{s=1}^{\infty} \sum_{i=1}^2  s^2 F^{ii}.
\end{align}
Note that we can rewrite the first of these as:
\begin{align}
	U_m &= 3 C \gamma_\perp^{2m-4} \gamma^{-m} u_\perp^2 I_m. \label{eq:UmFromIm}
\end{align}

Now, to handle the Bessel function sums, we do a second substitution:
\begin{align}
	\beta' = \frac{\beta_\perp}{(1-\beta_\parallel^2)^{1/2}} = \frac{u_\perp}{\gamma_\perp}. \label{eq:betaPerp}
\end{align}
Under this substitution, we have:
\begin{align}
	F^{11} &= \left(\frac{\cos \theta'}{\beta' \sin \theta' }\right)^2 \left[J_s\left(s \beta' \sin \theta'\right)\right]^2\\
	F^{22} &=  \left[J_s'\left(s \beta' \sin \theta'\right)\right]^2.
\end{align}

Now, the first thing to note is that the $F^{ii}$ are \emph{even} functions of $\cos \theta$. Thus, only the even part of $(1+\beta_\parallel \cos \theta')^{1-m}$ contributes, and we have:
\begin{alignat}{1}
	I_{m\in\{0,1\}} &= \int_{0}^{1} d w  \sum_{s=1}^{\infty} \sum_{i=1}^2  s^2 F^{ii} \\
	I_2 &= \int_{0}^{1} d w  (1-\beta_\parallel^2 w^2 )^{-1} \sum_{s=1}^{\infty} \sum_{i=1}^2  s^2 F^{ii} ,
\end{alignat}
with
\begin{align}
	F^{11} &= \left(\frac{w}{\beta' \sqrt{1-w^2}}\right)^2 \left[J_s\left(s \beta' \sqrt{1-w^2}\right)\right]^2\\
	F^{22} &=  \left[J_s'\left(s \beta' \sqrt{1-w^2}\right)\right]^2.
\end{align}
where we made the substitution $w \equiv \cos \theta'$.

Now, two useful sum identities (the Schott series, according to Melrose) exist here for Bessel functions:
\begin{align}
	\sum_{s=1}^{\infty} s^2 J_s^2 (s x) &= \frac{x^2 (4+x^2)}{16(1-x^2)^{7/2}}\\
	\sum_{s=1}^{\infty} s^2 J_s'^{\,2} (s x) &= \frac{4+3 x^2}{16(1-x^2)^{5/2}}.
\end{align}
With these sums, we find:
\begin{align}
	\sum_s s^2 F^{11} &= u^2 \frac{4 + \beta'^2 (1-u^2)}{16 (1-\beta'^2 (1-u^2))^{7/2}}\\
	\sum_s s^2 F^{22} &= \frac{4 + 3\beta'^2 (1-u^2)}{16 (1-\beta'^2 (1-u^2))^{5/2}}.
\end{align}

The integral evaluates nicely for $I_0 = I_1$:
\begin{align}
	I_0 = \frac{1}{3} \lp 1-\beta'^2 \rp^{-2} = \frac{\gamma_\perp^4}{3}, \label{eq:I0}
\end{align}
where in the last lines we used $1-\beta'^2 = \gamma_\perp^2$.
Using the additional identities $(1-\beta_\parallel^2) = \gamma_\perp^2/\gamma^2$ and $\beta_\perp = u_\perp / \gamma$, we can write:
\begin{align}
	U_{m\in\{0,1\}} = C \gamma_\perp^{2m} \gamma^{-m} u_\perp^2, \label{eq:U01}
\end{align}
recovering Bornatici's Eq.~(21).

Now, using Eq.~(\ref{eq:I0}) and Eq.~(\ref{eq:UmFromIm}), we can write $U_2$ in the form of Eq.~(\ref{eq:U2FromDelta}):
\begin{align}
	U_2 = C \gamma_\perp^{4} \gamma^{-2} u_\perp^2 \left(1 + \frac{u_\parallel^2}{u_\perp^2} \Delta \right),
\end{align}
where
\begin{align}
	\Delta = \frac{u_\perp^2}{u_\parallel^2} \left(\frac{I_2}{I_0} - 1\right).
\end{align}
The integral $I_2$ also evaluates (thanks to Mathematica), but lacks the aesthetic minimalism of its peers $I_0$ and $I_1$, giving Eq.~(\ref{eq:DeltaFull}).

\section{Nonrelativistic Limit of $\lambda$} \label{app:NonrelativisticLambda}

Following Eq.~(\ref{eq:EmmissionSumS}), the radiation emitted in a harmonic $s$ is proportional to:
\begin{alignat}{1}
	P_{s} &\equiv  \int_{0}^{1} d w  \sum_{i=1}^2  s^2 F^{ii}, 
\end{alignat}
Following Bekefi,\cite{BekefiRadiationProcesses} we ignore the O mode ($F_{11}$).
For the X mode, we have:
\begin{align}
	F^{22} &=  \left[J_s'\left(s \beta' \sqrt{1-w^2}\right)\right]^2.
\end{align}
In the nonrelativistic limit, $\beta' \ll 1$, and we can expand the $J_s'(s x)$ in small argument as:
\begin{align}
	J_s'(s x) \approx \frac{1}{2 \Gamma(s)} \lp \frac{s x}{2} \rp^s. 
\end{align}
Thus, the power radiated from mode $s$ becomes:
\begin{alignat}{1}
	P_{s} &=  \frac{s^{2s} \lp \beta'\rp^{2s-2}}{2^{2s} \Gamma(s)^2} \int_{0}^{1} d w \lp 1-w^2\rp^{s-1} F^{ii}\\
	&= \frac{\sqrt{\pi}}{2} \frac{s^{2s} \lp \beta'\rp^{2s-2}}{2^{2s} \Gamma(s)^2} \frac{\Gamma(s)}{\Gamma(s+\tfrac{1}{2})}
\end{alignat}

Using the Legendre duplication formula, this becomes:
\begin{alignat}{1}
	P_{s} &= \frac{1}{4} \frac{s^{2s} \lp \beta'\rp^{2s-2}}{\Gamma(2s)}
\end{alignat}

Using Stirling's approximation:
\begin{align}
	\Gamma(x) \approx \sqrt{2 \pi (x-1)} \lp \frac{x-1}{e} \rp^{2s-1},
\end{align} 
and the approximation:
\begin{align}
	\lp \frac{s}{s-\tfrac{1}{2}} \rp^{2s-\tfrac{1}{2}} \approx e,
\end{align}
we eventually arrive at:
\begin{alignat}{1}
	P_{s} &= \frac{1}{4 \sqrt{\pi} \lp \beta'\rp^2} s^{1/2} \left(\frac{e\beta'}{2}\right)^{2s}.
\end{alignat}
Using the definition of $\beta'$, we arrive at:
\begin{alignat}{1}
	P_{s} &= \frac{1}{4 \sqrt{\pi} \lp \beta'\rp^2} s^{1/2} \left(\frac{e u_\perp}{2 \gamma_\perp}\right)^{2s}.
\end{alignat}

Because of the strong decline in the emission with increasing $s$ at small $u_\perp$, the approximate radiation escape fraction
\begin{align}
	\lambda \approx \frac{\sum_{s=m^*}^{\infty} P_s}{\sum_{s=1}^{\infty} \int_{0}^{1} d w P_s} .
\end{align}
can be approximated by the first term in each series:
\begin{align}
	\lambda \approx \frac{P_{m^*}}{P_1} = s^{1/2} \left(\frac{e u_\perp}{2 \gamma_\perp}\right)^{2s}. \label{eq:LambdaAsymptoticLowU}
\end{align}

\begin{figure}[h]
	\centering
	\includegraphics[width=\linewidth]{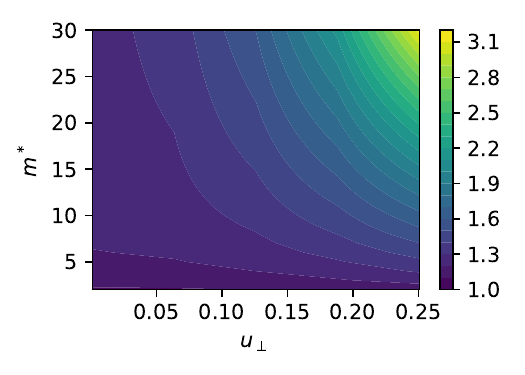}
	\caption{Ratio between asymptotic approximation to $\lambda$ as $u_\perp \rightarrow 0$ [Eq.~(\ref{eq:LambdaAsymptoticLowU})], and the full value as calculated from Eq.~(\ref{eq:LambdaGeneral}).}
	\label{fig:LowULambda}
\end{figure}

The approximation can be numerically checked against the full calculation of $\lambda$ from the main text; the ratio between the approximate and true quantities is shown in Fig.~\ref{fig:LowULambda}.
Of course, the approximation diverges rapidly as $u_\perp / \gamma_\perp \rightarrow 2/e$, but it holds well for very low values of $u_\perp$.

\section{Conversion to Fokker-Planck Coordinates} \label{app:CoordinateConversion}

To use the Fokker-Planck code from Ref.~\onlinecite{Ochs2023ConfinementTime}, we must go from coordinates $(u_\perp,u_\parallel)$ to $(x,\theta)$, where
\begin{align}
	x &\equiv \frac{p}{p_{th}} = \frac{m_e c u}{\sqrt{2 m_e T_0}} = \frac{u}{\sqrt{2 \chi_0}}\\
	\theta &= \arccos\lp\frac{u_\parallel}{u}\rp
\end{align}
with the inverse transform:
\begin{align}
	u_{\perp} &= \sqrt{2 \chi_0} x \sin \theta \\
	u_\parallel &= \sqrt{2 \chi_0} x \cos \theta.
\end{align}

Using standard tensor transformations to convert the diffusion coefficients, we find that the $(x,\theta)$ form of the diffusion operator is given by:
\begin{align}
	\Gammav &= \frac{1}{\sqrt{2 \chi_0}} \bvec \Gamma_\parallel \cos \theta + \Gamma_\perp \sin \theta \\   (\Gamma_\perp \cos \theta - \Gamma_\parallel \sin \theta)/x \evec \\
	\Dv &= \frac{1}{2\chi_0} \bvec D^{xx} & D^{x\theta}\\ D^{x\theta} & D^{\theta\theta}\evec.
\end{align}
\begin{align}
	D^{xx} &= D^{\parallel \parallel} \cos^2 \theta +D^{\perp \perp} \sin^2 \theta + D^{\perp \parallel} \sin 2\theta \\
	D^{\theta \theta} &= \frac{1}{x^2} \lp D^{\perp \perp} \cos^2 \theta +D^{\parallel \parallel} \sin^2 \theta - D^{\perp \parallel} \sin 2\theta  \rp\\
	D^{x \theta} &= \frac{1}{x} \lp \frac{1}{2}\lp D^{\perp \perp} - D^{\parallel \parallel} \rp \sin 2 \theta + D^{\perp \parallel}\cos(2\theta) \rp.
\end{align}

Plugging in the form of the diffusion operator, it can be put in the form of Eq.~(\ref{eq:DiffusionMatrixDecomposition}):
\begin{align}
	\Gammav &= -\nu_R \ve{g} \\	
	\ve{D} &= \nu_R \frac{T_\text{bb}}{m c^2} \frac{\gamma_\perp^2}{\gamma} \frac{1}{u_\perp^2} (1-\lambda) \left(\ve{g}\ve{g} + \Delta \ve{h}\ve{h} \right),
\end{align}
but now with
\begin{align}
	\ve{g} &= \gamma_\perp^{-2} \bvec \gamma^2 x \sin^2 \theta \\ \cos\theta \sin \theta \evec ; \quad \ve{h} = \bvec 0 \\ 1 \evec.
\end{align}

In these coordinates, the metric which appears in Eq.~(\ref{eq:FundamentalDiffusion}) becomes:
\begin{align}
	\sqrt{g} &= x^2 \sin \theta.
\end{align}

%


\clearpage\newpage

\end{document}